\documentclass[preprint,times,authoryear]{elsarticle}
\usepackage{amssymb}
\usepackage{lipsum}
\usepackage{amsmath}
\usepackage[colorlinks=true,citecolor=blue,pdfauthor=author]{hyperref}
\usepackage{graphicx,graphics,color}
\usepackage{ulem}
\usepackage{color}
\usepackage{lineno}
\usepackage[T1]{fontenc}
\usepackage{lscape}
\usepackage{threeparttable}

\newcommand{\arcsec}{\ensuremath{^{\prime\prime}}}

\journal{New Astronomy}

\begin{document}

\begin{frontmatter}

\title{The influence of AGN feedback on star formation in red spiral galaxies}

\author[a]{Moreom Akter}
\ead{ma25u@fsu.edu}

\author[a]{Wayne A. Barkhouse\corref{cor1}}
\ead{wayne.barkhouse@und.edu}
\cortext[cor1]{Corresponding author}
\affiliation[a]{organization={Department of Physics and Astrophysics, University of North Dakota},
            addressline={101 Cornell Street}, 
            city={Grand Forks},
            postcode={58202}, 
            state={North Dakota},
            country={USA}}

\author[b]{Sandanuwan P Kalawila Vithanage}
\ead{skalawil@gettysburg.edu}
\affiliation[b]{organization={Department of Physics, Gettysburg College},
            city={Gettysburg},
            postcode={17325},
            state={Pennsylvania},
            country={USA}}

\author[c]{Gihan L. Gamage}
\ead{glgamage@nmsu.edu}
\affiliation[c]{organization={Arts and Science Division, New Mexico State University - Alamogordo},
            city={Alamogordo},
            postcode={88310},
            state={New Mexico},
            country={USA}}

\author[d]{Omar L\'opez-Cruz}
\ead{omarlx@inaoep.mx}
\affiliation[d]{organization={INAOE},
            city={Tonantzintla},
            postcode={72840},
            state={Puebla},
            country={Mexico}}

\begin{abstract}
We investigated the influence of Active Galactic Nuclei (AGN) feedback on star formation in red spiral galaxies by analyzing a sample of 324 red and 273 blue face-on spirals selected from 115 low-redshift galaxy clusters. This multi-wavelength dataset combines optical emission line data from the Sloan Digital Sky Survey with X-ray fluxes from {\it Chandra} and {\it XMM-Newton} X-ray space telescopes. Using diagnostic emission line ratios, we constructed Baldwin-Phillips-Terlevich (BPT) diagrams, introduced in 1981 to classify galaxies based on nuclear activity. Our analysis reveals that most red spirals exhibit AGN or low-ionization nuclear emission-line region (LINER) signatures, based on optical data, as determined by their location in the BPT, Cid Fernandes, and Mazzolari diagrams. These results are consistent with the presence of negative feedback from AGNs quenching star formation. Conversely, X-ray analysis reveals that many blue spirals exhibit high X-ray luminosities and are situated in the AGN region of emission line ratio diagrams, suggesting that AGN-driven positive feedback may be enhancing star formation. Our findings support the hypothesis that AGN feedback plays a key role in the evolution of spiral galaxies, particularly in quenching star formation and driving the transition from blue to red spiral systems. 
\end{abstract}

\begin{keyword}
\texttt{}galaxies: active -- galaxies: nuclei -- galaxies: spiral -- galaxies - star formation
\end{keyword}
\end{frontmatter}

\section{Introduction}
\label{sect:intro}
Active galactic nuclei (AGN) are among the most luminous and energetic phenomena in the Universe, powered by the accretion of matter and gas onto supermassive black holes (SMBHs) at the center of galaxies \citep[e.g., ][]{Shakura1973, Kormendy2013,Peterson2013}. These processes release energy across the electromagnetic spectrum \citep{Shang2011,Padovani2017}, and AGNs are recognized as drivers of galaxy evolution through feedback mechanisms that regulate star formation \citep[e.g.,][]{Bower2006,Fabian2012, Silk2013, Zinn2013, Dubois2016,Stemo2020,Yao2021}. Feedback can be either negative, quenching star formation by ejecting or heating gas \citep[e.g.,][]{Luo2021, Raouf2024, Rubinur2024}, or positive, triggering star formation by compressing interstellar matter \citep[e.g.,][]{Masters2010,Silk2013,Wagner2016,Morganti2017,Cresci2018,Abdulrahman2023,Venturi2023}.

Spiral galaxies are rich in gas and dust, which fuels the formation of young, massive, hot blue stars, making them appear brighter and bluer than most other galaxy types \citep{Singh2013}. Spiral galaxies provide a valuable framework for investigating AGN feedback across different evolutionary stages \citep{Bergh1998}. The placement of galaxies in the color-magnitude diagram (CMD) reveals a bimodal distribution; star-forming blue cloud, quiescent red-sequence, and the transitional green valley \citep[e.g., ][]{Kauffmann2003, Bell2004, LopezCruz2004, Faber2007, Salim2014}. Numerous studies have identified an enhanced frequency of AGNs within green valley galaxies, implying that AGN activity may play a direct role in the transitioning of galaxies from blue to red \citep[e.g.,][]{Pierce2007, Schawinski2007, Povic2012, Wang2017, Lacerda2020}.

The impact of AGNs on galaxy properties is closely tied to structural features such as bars, which are known to drive gas toward galaxy centers and potentially fuel SMBH accretion \citep[e.g.,][]{Diaz2020}. However, some studies \citep[e.g.,][]{Noguchi1996,Lesser2022} argue that bar instabilities alone are not sufficient to explain quenching in red spirals, reinforcing the role of AGNs as regulators of star formation (Lesser et al. 2026, in preparation). Additionally, LINERs \citep[low-ionization nuclear emission-line regions;][]{Heckman1980} are commonly found in the bulges of early-type spiral galaxies, which have a higher fraction of older stars compared to late-type spirals, and can contain low-luminosity AGNs (LLAGN) through shock-ionized gas \citep[e.g.,][]{Heckman1980, Ferland1983,Halpern1983,Terlevich1985,Ho2003,Kauffmann2003}.

Advances in multi-wavelength surveys have significantly improved our ability to identify and classify AGNs. Optical diagnostics such as the BPT diagram \citep[e.g.,][]{Baldwin1981,Kewley2001a,Kewley2001b,Kauffmann2003,Schawinski2007,Venturi2023,Mazzolari2024,Mazzolari2025} are now routinely complemented with infrared \citep{Jarrett2011,Jarrett2013}, ultraviolet \citep{Mahajan2018}, and X-ray diagnostics \citep{Brandt2015}. These multi-wavelength tools are essential for resolving ambiguities in AGN classifications, particularly in obscured systems or galaxies with composite spectra \citep[e.g.,][]{Hickox2018,Agostino2019,Birchall2020,Birchall2022}. 

Recent studies have emphasized that LINER classifications, particularly in the [N II]/H$\alpha$ BPT diagram, may not reliably distinguish AGN-powered emission from ionization due to old stellar populations or shocks \citep{Cheng2025}. Although [S II] and [O I] yield a better method to discriminate Seyferts and LINERs, the [N II] diagram lacks a unified boundary, leading to misclassification in AGN census studies. This highlights the need for revised diagnostic criteria, such as the recently proposed Seyfert–LINER demarcation line \citep{Cheng2025}, which improves the consistency of classification between BPT branches.

In this study, we investigate the impact of AGN feedback on star formation using a sample of 324 red and 273 blue spiral galaxies drawn from 115 low-redshift ($0.012 < z < 0.179$) galaxy clusters. Our goal is to test the hypothesis that AGNs can exert both negative and positive feedback on star formation, and thus play a dual role in shaping the evolution of spiral galaxies. In Section \ref{sec:data} we describe our data and sample selection. We present key results from optical emission-line ratio and X-ray analyses in Section~\ref{sec:result}. The discussion is given in Section~\ref{sec:discussion}, and in Section~\ref{sec:conclusion} we summarize the conclusions of this study.

We use the Friedmann–Lema{\^i}tre-Robertson–Walker cosmology with $H_0 = 70$ km s$^{-1}$ Mpc$^{-1}$, $\Omega_{\Lambda} = 0.7$, and $\Omega_{\text{M}} = 0.3$ throughout this study.

\section{Data and sample selection}\label{sec:data}

This study is based on a multi-wavelength dataset that includes Sloan Digital Sky Survey (SDSS) optical fiber spectroscopy and X-ray observations, thus providing a wavelength coverage sensitive to multiple emission mechanisms relevant to both nuclear and star-forming activity.

\subsection{Cluster membership and morphology}

Galaxies were selected from a well-defined sample of 115 low-redshift ($0.012 < z < 0.179$) clusters, 14 of which were observed with the 3.6-meter Canada-France-Hawaii Telescope (CFHT) using the MegaCam detector in the $u$- and $r$-passband, with data reduction, object detection, and galaxy photometry fully described in \citet{Rude2020}. In addition, $B$- and $R_{c}$-band observations of 31 clusters were obtained with the KPNO 0.9-meter telescope using the T2KA CCD detector and the MOSAIC 8K camera \citep[see][for details regarding object detection and photometry]{LopezCruz2004,Barkhouse2007,Barkhouse2009}. Supplementary archival imaging data and catalogs were included from the WINGS survey \citep{Fasano2006,Varela2009,Valentinuzzi2011}, providing $B$- and $V$-band observations of 70 clusters obtained with the 2.5-meter INT and the MPG/ESO 2.2-meter telescopes \citep[see][for details regarding catalog and image access]{Moretti2014}.

All galaxy cluster catalogs from the 115-cluster sample have been analyzed in a homogeneous fashion using the same procedure to measure red-sequence slopes, y-intercepts, and dispersions from cluster color-magnitude diagrams \citep[see][for details regarding the fitting method]{LopezCruz2004}. The faint-end magnitude cut-off for each cluster is based on the limiting magnitude in the $r$/$R_{c}$/$V$-band, where the mean $2.5\sigma$ color uncertainty in the cluster red-sequence exceeds the $\pm 3\sigma_{d}$ dispersion of the red-sequence. The red-sequence dispersion ($\sigma_{d}$) is measured by a Gaussian fit to the color-magnitude distribution \citep{LopezCruz2004,Barkhouse2007}. All galaxies fainter than the limiting magnitude of the host cluster are excluded from further analysis. 

To minimize selection effects, red spirals were defined as those that are located within $\pm 3\sigma_{d}$ of the red-sequence. Blue spirals are those that are found $> 3\sigma_{d}$ blueward of the red-sequence in the color-magnitude diagram. Spectroscopic redshifts were obtained from SDSS and NASA/IPAC Extragalactic Database (NED) to confirm cluster membership and minimize inclusion of interlopers. Galaxies were included in our sample if their redshifts are within $\pm3\sigma_{\text{v}}$ of the cluster velocity dispersion \citep{Yahil1977} relative to the brightest cluster galaxy (BCG) of the host cluster.

A k-correction was applied to each galaxy based on redshift and color, and was calculated from \citet{Chilingarian2010} and \citet{Chilingarian2012}. We used $R_{V}=3.1$ for Milky Way extinction corrections \citep{Schlafly2011}, and no correction was made for host galaxy extinction. Only face-on or nearly face-on spirals with obvious spiral arm structures were included to minimize the effect of dust extinction and to preclude the inclusion of edge-on S0 galaxies. Spiral galaxies were selected via visual verification to avoid the inclusion of non-spiral systems. 

Our final sample of face-on cluster spiral galaxies consists of 273 blue spirals and 324 red spirals. Based on SDSS DR 17 photometry, extinction- and k-corrected absolute magnitudes $M_{r}$ of blue spirals range from $-17.1> M_{r} > -23.0$, with an average of $-21.2$, while red spirals range from $-17.3 > M_{r} > -23.3$, with a mean of $-21.2$. The average redshift of our sample of blue spirals is $z=0.076$, while the average for red spiral galaxies is $z=0.067$. 

Using the relationship between $M_{r}$ and galaxy mass given by \citet[][see their Figure 2]{Mahajan2018}, the mass of blue spirals range from $2.4\times 10^{8} M_{\odot}$ to $1.3\times 10^{11} M_{\odot}$. The mass range of red spirals extends from $3.0\times 10^{8} M_{\odot}$ to $1.8\times 10^{11} M_{\odot}$.

\subsection{SDSS optical spectroscopy}

Optical fiber spectroscopy data were obtained from the MPA-JHU and SDSS DR18 value-added catalogs \citep[for details, see][]{York2000,Brinchmann2004,Tremonti2004,Almeida2023} using standard S/N $\geq 3$ cuts for reliability. MPA-JHU spectral measurements have been corrected for galactic extinction and reddening. Emission-line fluxes were extracted for [O III] $\lambda\lambda$5007~\AA, 4363~\AA; [N II] $\lambda$6584~\AA; [S II] $\lambda\lambda$6717~\AA, 6731~\AA; [O I] $\lambda$6300~\AA; [O II] $\lambda\lambda$3726~\AA, 3729~\AA; [Ne III] $\lambda$3869~\AA; [He II] $\lambda$4686~\AA, H$\alpha$, H$\beta$, and H$\gamma$. These measurements were used to construct emission-line diagnostic diagrams for classifying AGN and star-forming galaxies. Although SDSS galaxy spectroscopy is obtained using 3\arcsec diameter fibers centered on each measured galaxy, this ``aperture bias'' is expected to have minimal impact on our results since we are probing the contribution to star formation from AGNs that are located in the central regions of galaxies. For the redshift range of our sample ($0.012 < z < 0.179$), a 3\arcsec diameter fiber corresponds to a physical size of 0.8 kpc to 9.0 kpc.

\subsection{{\it Chandra} and {\it XMM-Newton} X-ray data}

Publicly available X-ray data from the {\it Chandra X-ray Observatory} Source Catalog Release 2 Series \citep[https://doi.org/10.25574/csc2;][]{Evans2024} and the {\it XMM-Newton} X-ray space telescope archive were matched to galaxy positions using a 2\arcsec search radius. When multiple matches were found, the closest match in position was used. Luminosities were calculated from fluxes in the 0.1–10 keV ({\it Chandra}) and 0.2–12 keV ({\it XMM-Newton}) bands using $L_X = 4\pi D_L^2 F (1+z)^{\gamma - 2}$, where $D_L$ is the luminosity distance, $F$ is the observed flux, $z$ is the redshift, and $\gamma = 1.7$ \citep[based on][]{Lehmer2016,Birchall2022}. When a galaxy was matched in both {\it Chandra} and {\it XMM-Newton} catalogs, the flux measurement with the smallest uncertainty was retained. This occurred for 12 out of 45 galaxies, with a difference in flux exceeding $3\sigma$ for two of the 12 galaxies ($10.2\sigma$ and $5.7\sigma$, see Appendix A, Figure~\ref{fig:A1}). 

Matching our sample to X-ray data yielded a detection for 23 (7.1\%) red spirals and 22 (8\%) blue spirals. Non-detections of the remaining spirals (552 out of 597, 92.5\%) was due to both a lack of coverage by {\it Chandra} and {\it XMM-Newton} telescopes (239 spirals out of 597, 40\%), and coverage but no detection due to sensitivity limits (313 out of 597, 52.4\%) . For red spirals, X-ray luminosities range from $4.342\times 10^{39}$ erg s$^{-1}$ to $9.950\times 10^{42}$ erg s$^{-1}$, with a mean of $5.796\times 10^{41}$ erg s$^{-1}$ and a dispersion (rms) of $3.062\times 10^{41}$ erg s$^{-1}$. For blue spiral galaxies, X-ray luminosities range from $4.139\times 10^{39}$ erg s$^{-1}$ to $5.056\times 10^{43}$ erg s$^{-1}$, with an average of $4.670\times 10^{42}$ erg s$^{-1}$ and a dispersion (rms) of $2.940\times 10^{42}$ erg s$^{-1}$.  

The detection of AGNs using X-ray data is usually based on selecting galaxies that have an X-ray luminosity above a specific threshold. For example, some studies \citep[e.g.,][]{Trouille2010,Trouille2011,CastelloMor2012,Pons2014} used an X-ray luminosity threshold of $L_X > 10^{42}$ erg s$^{-1}$. In contrast, \citet{Pons2016} used a low luminosity limit of $L_X > 10^{41}$ erg s$^{-1}$ based on their analysis of a subset of AGNs from the 12 Micron Galaxy Survey \citep{Brightman2011}. 

\begin{figure*}
\centering
\includegraphics[scale=1.0]{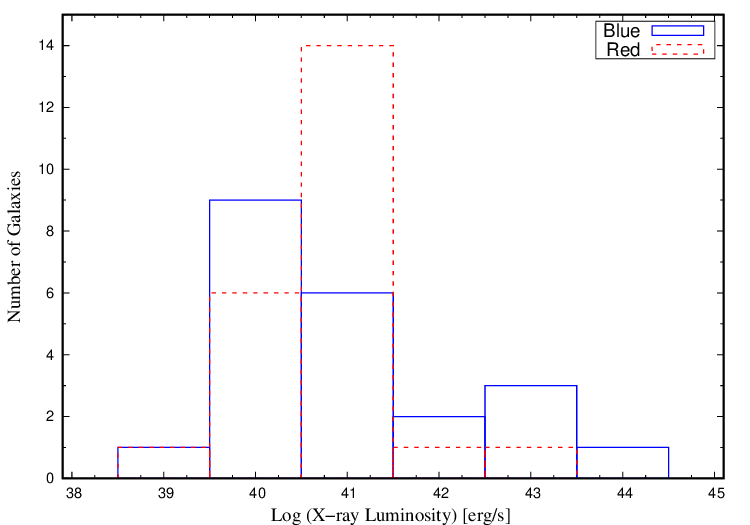}
\caption[Histogram of X-ray Luminosity Distribution]{X-ray luminosity histogram distributions for blue (solid boxes) and red cluster spiral galaxies (dashed boxes).}
\label{fig:redblue}
\end{figure*}

For our X-ray-detected galaxy sample, we found that 9 out of 22 blue spirals (40.9\%) have X-ray luminosities $> 10^{41}$ erg s$^{-1}$, with 22.7\% (5 out of 22) having luminosities $> 10^{42}$ erg s$^{-1}$. For the red spirals, 39.1\% (9 out of 23) have luminosities $>10^{41}$ erg s$^{-1}$ and 4.3\% (1 out of 23) have luminosities $> 10^{42}$ erg s$^{-1}$. Figure~\ref{fig:redblue} shows the distribution of X-ray luminosity for our sample of blue and red spiral galaxies.

%%%%%%%%%%%%%%%%%%%%%%
%%%%%%%%%%%%%%%%%%%%%%%%

\section{Results}\label{sec:result}

In this section, we describe the results of using optical emission-line diagnostic diagrams and X-ray luminosity analysis to evaluate the effects of possible AGN feedback on star formation in red and blue spiral galaxies located in low-redshift clusters.

\subsection{[O~III]/H$\beta$ vs [N~II]/H$\alpha$}
Optical emission-line diagnostic diagrams are widely used to classify galaxies based on their dominant ionization source. One of the most established method is the use of the BPT diagram, which utilizes the ratios [O~III]/H$\beta$ versus [N~II]/H$\alpha$ to distinguish AGN from star-forming galaxies. This diagnostic diagram is particularly effective because AGNs typically exhibit an elevated [O~III]/H$\beta$ ratio due to the presence of high-energy photons ($>$35~eV) needed for [O~III] emission, which are more common in AGN continua than in stellar photoionization \citep{Baldwin1981}. 

Additionally, AGNs often have a stronger [N~II]/H$\alpha$ ratio compared to normal star-forming galaxies \citep{Veilleux1987}. The physical origin of this enhancement lies in the hardness of the AGN radiation field. While young O- and B-type stars in star-forming galaxies produce a relatively soft ionizing spectrum that is rapidly absorbed, AGNs generate a hard power-law continuum extending into the extreme-UV and X-ray regime \citep{Baldwin1981,Ho2008}. These high-energy photons penetrate deep into the surrounding interstellar medium, creating extended partially ionized zones where neutral hydrogen and N$^{+}$ ions coexist \citep{Ferland1983,Kewley2019}. In such regions, energetic electrons efficiently collisionally excite nitrogen ions, significantly boosting the [N~II]~$\lambda6584$ emission relative to the recombination-dominated H$\alpha$ line. This effect can be further enhanced by shocks driven by AGN outflows or jets, which heat and compress the gas and increase the efficiency of low-ionization line emission \citep{Heckman1980,Rich2011,Farage2010}. In contrast, H II regions dominated by stellar photoionization have much thinner partially ionized zones and lack strong shock excitation, resulting in the characteristically lower [N~II]/H$\alpha$ ratios observed in star-forming galaxies.

Figure~\ref{fig:2} shows the distribution of galaxies in the standard BPT diagram using the emission-line ratios [O~III]/H$\beta$ versus [N~II]/H$\alpha$, enabling classification into star-forming, AGN/Seyfert, and LINERs. We note that we use ``AGN/Seyfert'' in our figures to avoid specific AGN subtype classifications. The black curve and straight line are from \citet{Kauffmann2003}, which separates star-forming, AGN/Seyfert, and LINER regions. For the star-forming region, we find that 79.4\% of blue galaxies (81 out of 102) occupy this region, while only 15.5\% of red galaxies (11 out of 71) are found in this area. In the AGN/Seyfert zone, 18.6\% (19) of blue galaxies and 64.8\% (46) of red spirals are located in this area. Lastly, in the LINER region we find 2\% (2) of blue spirals and 19.7\% (14) of red galaxies occupy this area. The number and percent of blue and red galaxies that occupy specific regions of the diagram are tabulated in Table~\ref{tab:galaxy_classification}, along with the results for other emission-line diagnostic diagrams. 

\subsection{[O~III]/H$\beta$ vs. [S~II]/H$\alpha$}
To further refine galaxy classification, we employed the Veilleux \& Osterbrock diagrams (hereafter VO87 diagrams; see, for example, Figure~\ref{fig:3}), which depict alternative line ratios: [O~III]/H$\beta$ vs. [S~II]/H$\alpha$ \citep{Veilleux1987}\footnote{[S~II] = (6717 + 6731)~\AA\; and [O~II] = (3726 + 3729)~\AA}. This diagram is especially useful for separating narrow-line AGNs from H~II region-like galaxies and are sensitive to LINERs, which are often associated with weak AGNs or shock excitation \citep[e.g.,][]{Heckman1980, Terlevich1985}.

For Figure~\ref{fig:3}, we find 93.4\% (85 out of 91) of blue galaxies are located in the star-forming region, while 3.3\% (3) are found each in the AGN/Seyfert and LINER sections. For red spirals, 27.8\% (15 out of 54) are in the star-forming area, while 5.6\% (3) are classified as AGN/Seyfert and 66.7\% (36) are LINER galaxies.

\subsection{[O~III]/H$\beta$ vs. [O~I]/H$\alpha$}

In Figure~\ref{fig:4}, we use the VO97 diagram of [O~III]/H$\beta$ vs. [O~I]/H$\alpha$ to help separate AGN/Seyfert galaxies from star-forming and LINER-like systems. For Figure~\ref{fig:4}, we find that 25 out of 31 blue spirals (80.6\%) are located in the star-forming region, 12.9\% (4) are found in the AGN/Seyfert zone, and 6.4\% (2) are in the LINER area. For the red spiral population, only 4.8\% (2 out of 42) are star-forming, 11.9\% (5) are classified as AGN/Seyfert, and 83.3\% (35) are in the LINER region.

In our sample, classification using the schemes of \citet{Dopita2000}, \citet{Kewley2001a}, \citet{Kewley2001b}, \citet{Kauffmann2003}, \citet{Kewley2006}, \citet{Mazzolari2024,Mazzolari2025}, and \citet{Scholtz2025} shows that most red spirals and a few blue spirals are identified as AGNs in the [N~II]/H$\alpha$ BPT diagram. Meanwhile, many of these same galaxies fall into the LINER regions in the VO87 diagrams, supporting the presence of hard ionizing sources such as AGNs or shocks. 

Conversely, blue spirals are predominantly distributed across the star-forming regions, suggesting that their ionization is mainly due to young stars, with minimal AGN contribution. Their lower [O~III]/H$\beta$ ratios indicate softer ionizing spectra, consistent with stellar photoionization. In contrast, red spirals exhibit systematically higher [O~III]/H$\beta$, [N~II]/H$\alpha$, and [S~II]/H$\alpha$ ratios, consistent with harder ionization fields and LINER-like excitation.

\subsection{[O~III]~$\lambda4363$/H$\gamma$ vs. [O~III]~$\lambda5007$/[O~II]}

We incorporated newer diagnostics proposed by \citet{Mazzolari2024,Mazzolari2025} and \citet{Backhaus2025}, including [O~III]~$\lambda4363$/H$\gamma$ vs. [O~III]~$\lambda5007$/[O~II]. Figure~\ref{fig:5} shows that many red spirals have high [Ne~III]/[O~II] and [O~III]/[O~II] ratios, indicative of AGN-like hard ionization. In Figure~\ref{fig:5}, 0\% of blue spirals (0 out of 10) are found in the ``star-forming or AGN/Seyfert'' region, while only 2.9\% (2 out of 68) of red spirals are location in this section of the diagram. Conversely, 100\% (10) of blue spirals and 97\% (66) of red galaxies are in the ``AGN/Seyfert-only'' section. 

\subsection{[O~III]~$\lambda4363$/H$\gamma$ vs. [Ne~III]/[O~II]}

An additional diagnostic plot proposed by \citet{Mazzolari2024,Mazzolari2025} and \citet{Backhaus2025} is the use of [O~III]~$\lambda4363$/H$\gamma$ vs. [Ne~III]/[O~II] to help discriminate between star-forming galaxies and AGN/Seyfert systems. Our data are plotted in Figure~\ref{fig:6}, where 11.1\% of blue spirals (1 out of 9) are in the ``star-forming'' zone, 11.1\% (1) in the ``SF or AGN/Seyfert'' section, and 77.8\% (7) are found in the ``AGN/Seyfert-only'' region. For red spirals, 3.6\% (2 out of 56 galaxies) are located in the ``star-forming'' area, 30.3\% (17) in the ``SF or AGN/Seyfert'' zone, and 66.1\% (37) in the ``AGN/Seyfert-only'' section. 

\subsection{[O~III]/[O~II] vs. [N~II]/H$\alpha$}

For sources with incomplete or low S/N in classical emission-line ratio diagrams, we adopted the WHAN-like\footnote{The WHAN diagram, introduced by \citet{Fernandes2010,Fernandes2011}, combines the H$\alpha$ equivalent width with the [N~II]/H$\alpha$ ratio to classify galaxies and to include weak-line systems that are excluded from classical BPT diagrams due to low H$\beta$ or [O~III] signal-to-noise.} extension [O~III]/[O~II] vs. [N~II]/H$\alpha$ as shown in Figure~\ref{fig:7} \citep{Fernandes2010}. This plot reveals AGN or LINER-like activity in galaxies misclassified by BPT due to weak lines or dust. In Figure~\ref{fig:7}, we find that 85 of 97 blue spirals (87.6\%) occupy the star-forming region of the figure, 8.2\% (8) are AGN/Seyfert, and 4.1\% (4) are classified as LINERs. For red spirals, 29.3\% (22 of 75 galaxies) are in the star-forming area, 32\% (24) are in the AGN/Seyfert zone, and 38.7\% (29) are located in the LINER region.

\subsection{[O~III]/H$\alpha$ vs. [N~II]/H$\alpha$}

An additional WHAN-like diagnostic plot uses [O~III]/H$\alpha$ vs. [N~II]/H$\alpha$ to help separate star-forming, AGN/Seyfert, and LINER systems (see Figure~\ref{fig:8}). Of the 95 blue galaxies plotted in Figure~\ref{fig:8}, 92.6\% (88 out of 95) are located in the star-forming region, 7.4\% (7) are AGN/Seyfert, and 0\% are LINER. For 62 red spirals, 24.2\% (15) are classified as star-forming, 59.7\% (37) are in the AGN/Seyfert region, and 16.1\% (10) are LINERs.

\subsection{Equivalent width of H$\alpha$ versus [N~II]/H$\alpha$}

To address galaxies with low S/N in H$\alpha$ or [N~II], we used the WHAN diagram shown in Figure~\ref{fig:9}, using the equivalent width (EW) of H$\alpha$ versus [N~II]/H$\alpha$, to distinguish between strong/weak AGNs, star-forming, retired, and passive galaxies \citep[see for details;][]{Fernandes2010,Fernandes2011}. Several red spirals classified ambiguously in BPT were confirmed as LINERs in WHAN, particularly those with H$\alpha$ EW < 3~\AA. Of the 155 blue spiral galaxies, 74.8\% (116) are ``pure star-forming,'' 14.2\% (22) are ``strong AGN,'' only 1.3\% (2) are ``weak AGN,'' 9.0\% (14) are ``retired galaxies,'' and 0.6\% (1) are classified as ``passive galaxies.'' Of the 114 red spiral galaxies, 15.8\% (18) are pure star-forming, 7.9\% (9) are strong AGN, 7.0\% (8) are weak AGN, 41.2\% (47) are retired galaxies, and 28.1\% (32) are passive galaxies.

Table~\ref{tab:galaxy_classification} presents a summary of galaxy classification based on the emission-line ratio diagnostic plots. Altogether, multiple emission-line diagnostics support that red spirals more frequently host AGNs or LINERs, while blue spirals are typically star-forming, with some showing composite characteristics suggesting coexistence with star-forming and AGN-like activity.

\subsection{X-ray emission and AGN identification}

To complement optical diagnostics, we examined X-ray luminosities from {\it Chandra} and {\it XMM-Newton}. A galaxy was classified as an X-ray AGN if its luminosity exceeded $10^{42}$ erg s$^{-1}$ \citep[e.g.,][]{Trouille2010,CastelloMor2012,Pons2016}, by which we could distinguish AGN-dominated sources from normal star-forming galaxies \citep[e.g.,][]{Fabbiano1989, Zezas1998, Moran1999}.

Among our X-ray-detected galaxy sample, 40.9\% of blue spirals (9 out of 22) and 39.1\% of red spirals (9 out of 23) exceed $10^{41}$ erg s$^{-1}$; 22.7\% of blue (5 out of 22) and 4.3\% of red spirals (1 out of 23) have $L_X > 10^{42}$ erg s$^{-1}$. This suggests AGNs are not exclusive to quiescent systems, supporting the dual-mode feedback model \citep[e.g.,][]{Cresci2018, Venturi2023}. In this framework, AGNs operate in two distinct modes based on the energy transfer mechanism. The higher-luminosity AGNs in our blue spiral sample correspond to the ``radiative mode'' (or quasar mode), where energy is released via radiation or winds from the accretion disk, ionizing and heating the surrounding gas \citep{Morganti2017}. This process regulates star formation by preventing gas from cooling or by expelling it through quasar-driven winds \citep{Harrison2014, Rubinur2024}. Conversely, the X-ray activity in our red spirals likely represents the ``kinetic mode'' (or radio mode), dominant in lower-power AGNs where mechanical energy is injected into the medium via jets and shocks \citep{Morganti2017}. The presence of significant X-ray luminosities across both populations indicates that AGN feedback is a multi-stage process, transitioning from radiative-dominated quenching in gas-rich blue systems to a kinetic-dominated mechanism in quiescent red systems.

\begin{figure*}
\centering
\includegraphics[scale=1.0]{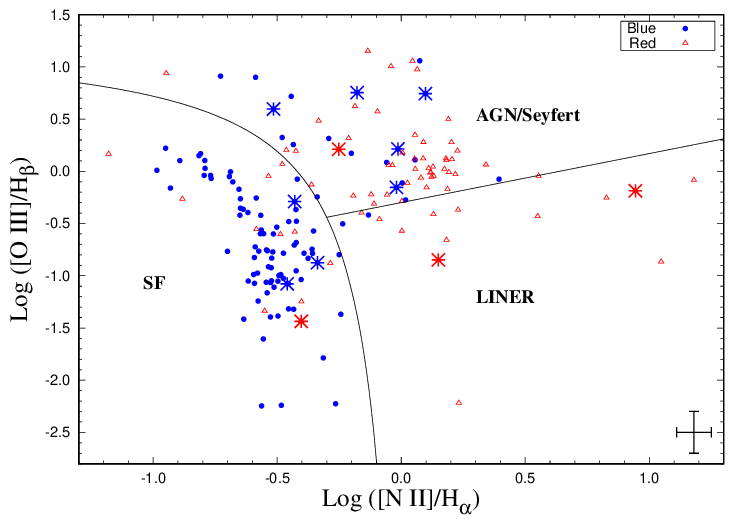}
\caption[BPT diagram]{Standard BPT diagram for log ([O III]/$H\beta$) vs. log ([N II]/$H\alpha$). Solid curve and straight line are from \citet{Kauffmann2003}, which separates AGN/Seyfert, star-forming, and LINER galaxies. Galaxies detected in the X-ray are depicted with asterisk symbols. The median 1$\sigma$ error bars are given in the lower-right corner of the figure.}
\label{fig:2}
\end{figure*}

\begin{figure*}
\centering
\includegraphics[scale=1.0]{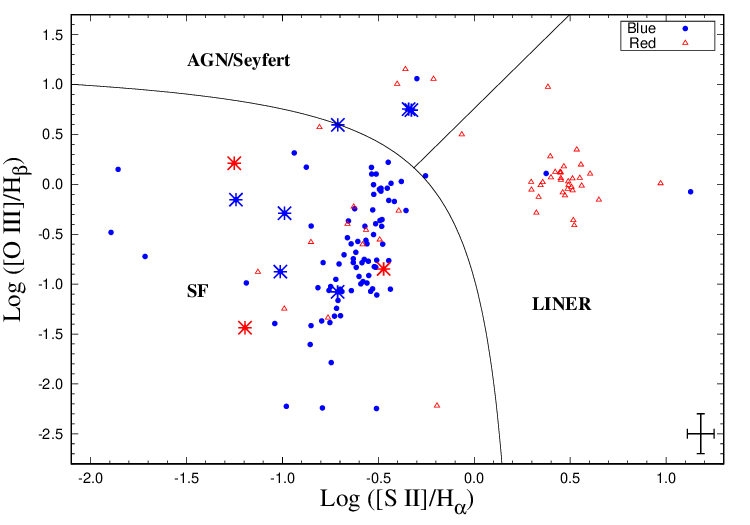}
\caption[First VO87 diagram]{VO87 diagram using log ([O III]/$H\beta$)  vs. log ([S II]/$H\alpha$). The black curve is from \citet{Kewley2001a,Kewley2001b}, which separate the star-forming (SF) region from the AGN/Seyfert area. The straight line, from \citet{Kewley2006}, depicts the demarcation line that separate AGN/Seyferts from LINER systems. Galaxies detected in the X-ray are represented by asterisk symbols. The median 1$\sigma$ error bars are given in the lower-right corner of the figure. }
\label{fig:3}
\end{figure*}

\begin{figure*}
\centering
\includegraphics[scale=1.0]{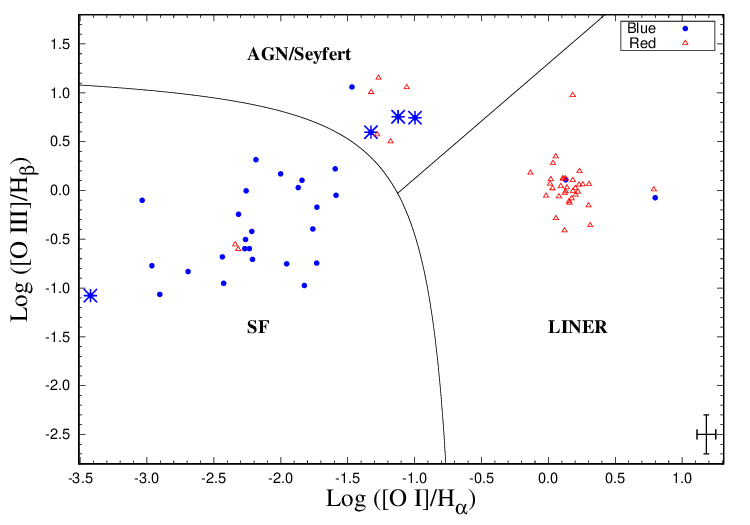}
\caption[second VO87 diagram]{VO87 diagram using log ([O III]/$H\beta$)  vs. log ([O I]/$H\alpha$). The black curve is from \citet{Kewley2001a,Kewley2001b}, which separate the star-forming (SF) region from the AGN/Seyfert area. The straight line, from \citet{Kewley2006}, depicts the demarcation line that separate AGN/Seyferts from LINER systems. Galaxies detected in the X-ray are represented by asterisk symbols. The median 1$\sigma$ error bars are given in the lower-right corner of the figure.}
\label{fig:4}
\end{figure*}

\begin{figure*}
\centering
\includegraphics[scale=1.0]{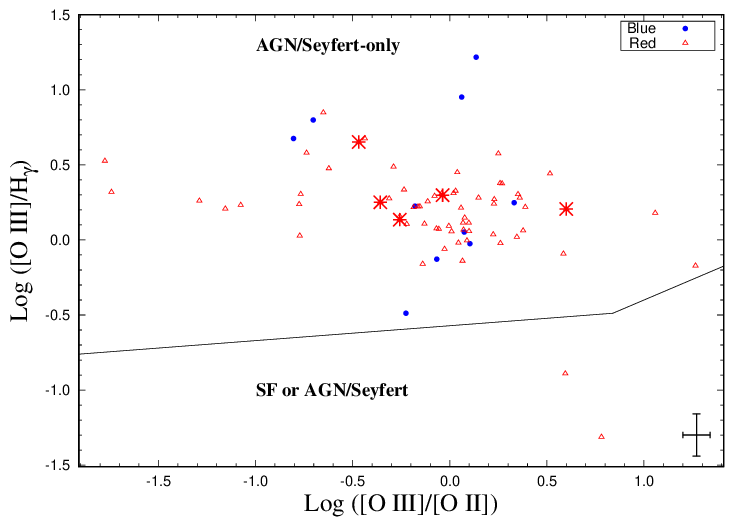}
\caption[Mazzolari diagram]{Mazzolari diagram \citep{Mazzolari2024} plotting log ([O III]/$H\gamma$)  vs. log ([O III]/[O II]). Straight line segments are from \citet{Mazzolari2024}, which separates galaxies into “AGN/Seyfert-only” and “SF or AGN/Seyfert” regions. Galaxies detected in the X-ray are represented by asterisk symbols. The median 1$\sigma$ error bars are given in the lower-right corner of the figure.}
\label{fig:5}
\end{figure*}

\begin{figure*}
\centering
\includegraphics[scale=1.0]{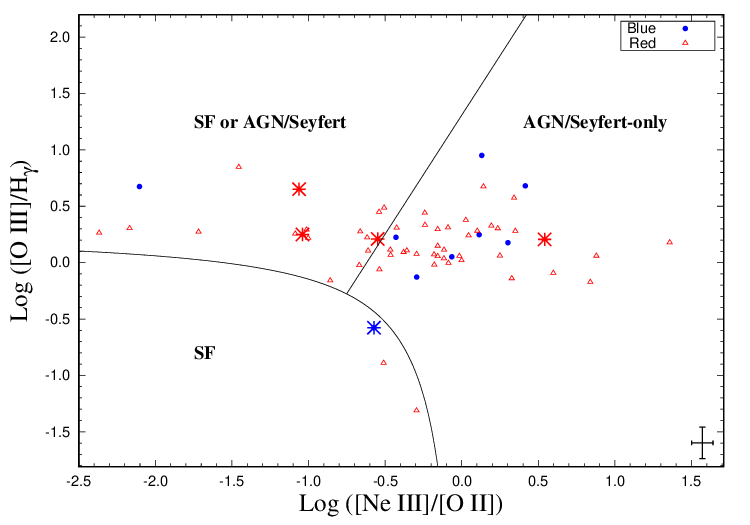}
\caption[Second Mazzolari diagram]{Mazzolari diagram \citep{Mazzolari2024} using log ([O III]/H$\gamma$) vs. log ([Ne III]/[O II]). The curve and straight lines are from \citet{Backhaus2025}, separating ``star-forming'' (SF), ``SF or AGN/Seyfert,'' and ``AGN/Seyfert-only'' regions. Galaxies detected in the X-ray are represented by asterisk symbols. The median 1$\sigma$ error bars are given in the lower-right corner of the figure.}
\label{fig:6}
\end{figure*}

\begin{figure*}
\centering
\includegraphics[scale=1.0]{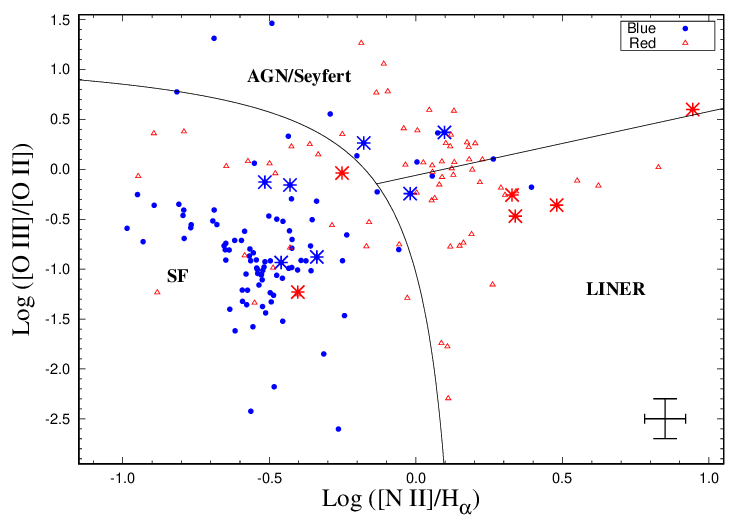}
\caption[Cid Fernandes diagram]{Cid Fernandes diagram \citep{Fernandes2010} using log ([O III]/[O II]) vs. log ([N II]/$H\alpha$). The black curve is from \citet{Kewley2001a,Kewley2001b}, which separate the star-forming (SF) region from the AGN/Seyfert area. The straight line, from \citet{Kewley2006}, depicts the demarcation line that separate AGN/Seyferts from LINER systems. Galaxies detected in the X-ray are represented by asterisk symbols. The median 1$\sigma$ error bars are given in the lower-right corner of the figure.}

\label{fig:7}
\end{figure*}

\begin{figure*}
\centering
\includegraphics[scale=1.0]{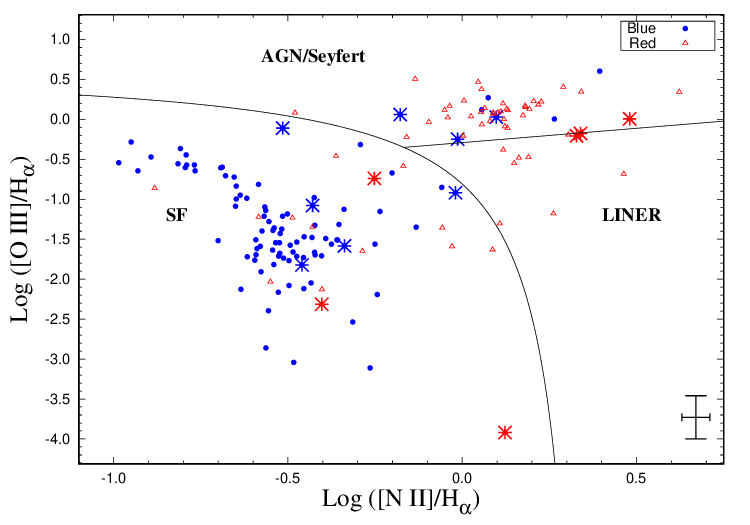}
\caption[second Cid Fernandes diagram]{Cid Fernandes diagram \citep{Fernandes2010} using log ([O III]/$H\alpha$) vs. log ([N II]/$H\alpha$). The black curve is from \citet{Kewley2001a,Kewley2001b}, which separate the star-forming (SF) region from the AGN/Seyfert area. The straight line, from \citet{Kewley2006}, depicts the demarcation line that separate AGN/Seyferts from LINER systems. Galaxies detected in the X-ray are represented by asterisk symbols. The median 1$\sigma$ error bars are given in the lower-right corner of the figure.} 
\label{fig:8}
\end{figure*}

\begin{figure*}
\centering
\includegraphics[scale=1.0]{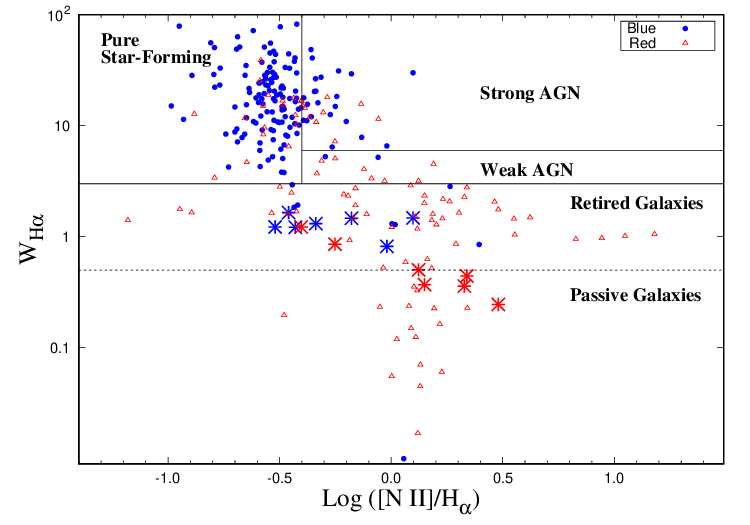}
\caption[WHAN diagram]{The WHAN diagram from \citet{Fernandes2011} using the equivalent width (EW) of H$\alpha$ ($W_{H\alpha}$) vs. log ([N~II]/H$\alpha$). The horizontal dashed line separates retired and passive galaxies, while the vertical and horizontal sold lines define regions containing pure star-forming, strong, and weak AGN. Galaxies detected in the X-ray are represented by asterisk symbols.}
\label{fig:9}
\end{figure*}

Only 35\% (113 out of 324) red spirals and 36\% (98 out of 273) blue spirals had measurable optical emission lines (see Figures~\ref{fig:2} to \ref{fig:9}), and most were LINERs or LLAGNs, with weak ionization features \citep{Kauffmann2003, Silverman2008}. The discrepancy between our X-ray detection rates and optical line measurements suggests that a significant fraction of AGN activity in these spirals is obscured or diluted by host galaxy light, particularly in massive, star-forming systems. Tables \ref{tab:4} and \ref{tab:5} gives the classification of galaxies based on X-ray luminosity and emission-line ratios.
 
Despite a broader range of X-ray luminosities in blue spirals, we performed a cumulative distribution comparison using the Kolmogorov-Smirnov (K–S) test based on measured $L_X$ values for red and blue galaxies. The test yielded a K–S statistic of $D=0.24$ and a probability of $P=49\%$, suggesting that the two galaxy populations are statistically consistent with being selected from the same parent distribution.
 
An important result is the partial mismatch between X-ray- and optically-selected AGNs, consistent with the findings by \citet{Agostino2019}, showing that optical diagnostics can miss AGNs in star-forming galaxies due to H~II contamination. This highlights the need for multi-wavelength AGN identification strategies.
%%%%%%%%%%%%%%%%%%%%%%
%%%%%%%%%%%%%%%%%%%%%%%%

%%%%%%%%%%%%%%%%%%%%%%%%%%%%%
%%%%%%%%%%%%%%%%%%%%%%%%%%%%%%%%

%%%%%%%%%%%%%%%%%%%%%%%%%%%%%
%%%%%%%%%%%%%%%%%%%%%%%%%%%%%%%%

%%%%%%%%%%%%%%%%%%%%%%%%%%%%%
%%%%%%%%%%%%%%%%%%%%%%%%%%%%%%%%

%%%%%%%%%%%%%%%%%%%%%%%%%%%%%
%%%%%%%%%%%%%%%%%%%%%%%%%%%%%%%%

%%%%%%%%%%%%%%%%%%%%%%%%%%%%%
%%%%%%%%%%%%%%%%%%%%%%%%%%%%%%%%

%%%%%%%%%%%%%%%%%%%%%%%%%%%%%
%%%%%%%%%%%%%%%%%%%%%%%%%%%%%%%%

%%%%%%%%%%%%%%%%%%%%%%%%%%%%%
%%%%%%%%%%%%%%%%%%%%%%%%%%%%%%%%

%%%%%%%%%%%%%%%%%%%%%%
\begin{figure*}
\centering
\includegraphics[scale=1.0]{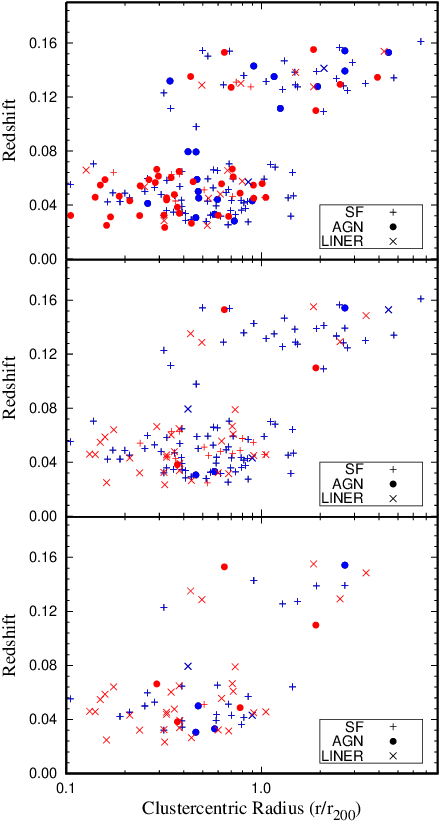}
\caption[Radial distribution of star-forming, AGN, and LINER-type systems]{{\it Top panel:} Redshift versus cluster-centric radius ($r/r_{200}$) for red/blue galaxies classified as ``star-forming,'' ``AGN,'' and ``LINER'' systems based on the standard BPT emission line ratio plot using [O III]/$H\beta$ vs. [N II]/$H\alpha$ (see Figure~\ref{fig:2}). {\it Middle panel:} Galaxy classification into ``star-forming,'' ``AGN,'' and ``LINER'' systems based on the BPT plot using [O III]/$H\beta$ vs. [S II]/$H\alpha$ (see Figure~\ref{fig:3}). {\it Bottom panel:} Galaxy classification into ``star-forming,'' ``AGN,'' and ``LINER'' systems based on the BPT plot using [O III]/$H\beta$ vs. [O I]/$H\alpha$ (see Figure~\ref{fig:4}). For all three panels, galaxies classified as ``star-forming'' are depicted by plus symbols, ``AGNs'' are denoted by solid circles, and ``LINER'' galaxies are represented by cross symbols.}
\label{fig:10}
\end{figure*}
%%%%%%%%%%%%%%%%%%%%%%%%

Although detailed {\it WISE}-based infrared AGN classification is presented in Paper I (Section 3.3), we include Table~\ref{tab:wise_agn} here to summarize the X-ray properties of galaxies identified as AGNs in the {\it WISE} color-color diagram for our sample.

\subsection{Statistical verification of emission-line distributions}

K–S tests were performed to determine whether red spirals are affected more by the presence of AGN compared to blue spirals by measuring the statistical significance that both groups of spirals have the same parent distribution using various emission-line ratios. The results tabulated in Table~\ref{tab:ks_test} indicate that based on five of the seven emission-line ratio plots, blue and red spirals are statistically not selected from the same parent population. Only Figure~\ref{fig:5} based on log ([O III]/$H\gamma$)  vs. log ([O III]/[O II]) and Figure~\ref{fig:6} using log([O III]/H$\gamma$) vs. log ([Ne III]/[O II]), indicate that blue and red spirals have properties consistent with selection from the same parent distribution.   

\section{Discussion}\label{sec:discussion}

Our findings are consistent with the dual role of AGN feedback in spiral galaxies. Optical emission-line diagnostics reveal that a significant fraction (71\% to 97\%) of red spirals exhibit AGN and LINER signatures, consistent with negative feedback processes that suppress star formation and contribute to quenching. This aligns with previous studies \citep[e.g.][] {Schawinski2007, Fernandes2011,Povic2012}, which argue that AGNs are instrumental in transitioning galaxies across the green valley toward the red sequence.

On the other hand, there is a noticeable difference between optical and X-ray AGN classifications for blue spirals. Many optically classified star-forming or composite (star-forming/AGN) galaxies exhibit high X-ray luminosities (log $L_{X}$ > $10^{41}$ erg s$^{-1}$), suggesting hidden or diluted AGN activity not detectable in the optical regime due to ongoing star formation. This supports findings by both \citet{Hickox2009} and \citet{Birchall2022}, who highlight the need for multi-wavelength approaches to fully capture the AGN population, particularly in blue galaxies where star formation masks traditional signatures.

The combination of BPT, WHAN, and alternative diagnostic diagrams \citep[e.g.,][]{Kewley2001a,Kewley2001b,Kewley2006,Mazzolari2024,Mazzolari2025,Backhaus2025} enhanced the robustness of galaxy classification, revealing the shortcomings of relying solely on a single method. In particular, the WHAN diagram proved valuable in identifying weak or retired AGNs, especially among LINER-classified systems where ionization may arise from post-AGB stars or shocks \citep[e.g.][]{Fernandes2011}.

In Figure~\ref{fig:10}, we show the redshift vs. cluster-centric radius ($r/r_{200}$) of red and blue galaxies classified as ``star-forming,'' ``AGN,'' and ``LINER'' systems. In the top panel of Figure~\ref{fig:10}, galaxy classification is based on the standard BPT plot using [O III]/$H\beta$ vs. [N II]/$H\alpha$ (see Figure~\ref{fig:2}). In the middle panel, the [O III]/$H\beta$ vs. [S II]/$H\alpha$ emission line ratios indicate activity (see Figure~\ref{fig:3}). Finally, the bottom panel depicts galaxies classified based on activity indicated by [O III]/$H\beta$ vs. [O I]/$H\alpha$ (see Figure~\ref{fig:4}). For all three panels, projected cluster-centric distance was normalized to $r_{200}$ (i.e., $r/r_{200}$), the radius of a sphere whose density is 200 times the critical density of the universe \citep[cf.][]{Barkhouse2007}.

Examination of Figure~\ref{fig:10} indicates that red spirals that occupy a radius $(r/r_{200}) < 2$, and hence are most likely older cluster members compared to blue spirals at larger cluster-centric distance, have a higher fraction of AGN. Combining galaxy classifications from all three panels in Figure~\ref{fig:10}, we find that 32\% of red spirals with $(r/r_{200}) < 2$ are AGN, while only 11\% of blue spirals selected from within the same radius are AGN. This may be a result of the impact of AGN feedback in helping to transition blue spirals into red systems in galaxy cluster environments  \citep{Martini2002,Lokas2022}.

For a complementary analysis of stellar population age and star formation activity, including D$_n$(4000) and specific star formation rate metrics, we refer to Paper I \citep{Barkhouse2025} which explores the UV and IR properties of our galaxy sample. The D$_n$(4000) metric \citep{Balogh1999} is a spectral line index constructed from the ratio of the red (measured between 4000 and 4100 {\AA}) to the blue continuum (3850 to 3950 {\AA}). This ratio is sensitive to stellar age and star formation such that a larger value of  D$_n$(4000) implies a greater stellar population age and a lower star formation rate. The star formation rate is obtained from H$\alpha$ spectral line measurements and converted to luminosity using galaxy redshift information. The star formation rate is given by SFR~$(M_{\odot}\,yr^{-1})=7.93\times 10^{-42}~\text{L}\, (H\alpha)$, where the H$\alpha$ luminosity, L~(H$\alpha$), is in erg~s$^{-1}$ \citep{Bell2001,Dhiwar2023}. The specific star formation rate, the star formation rate normalized to galaxy mass, is obtained by dividing the SFR by mass estimated from the extinction- and k-corrected $M_{r}$ absolute magnitude of each galaxy \citep{Mahajan2018}.

These results contribute to the growing evidence that AGNs not only quench but may also stimulate star formation under specific conditions, reinforcing the idea of ``positive feedback'' \citep[e.g.][]{Wagner2016, Venturi2023}. In this study, galaxy environments likely amplify these effects by influencing gas availability and dynamical interactions.

\section{Conclusions}\label{sec:conclusion}

This study investigates the relationship between AGN feedback and star formation in 597 spiral galaxies from 115 low-redshift galaxy clusters, using a multi-wavelength dataset incorporating optical and X-ray observations. Key findings include:

\begin{itemize}
    \item A majority of red spiral galaxies (70.7\% to 97.0\%, depending on the diagnostic, see Table~\ref{tab:galaxy_classification}) display AGN or LINER signatures in optical diagnostics, consistent with feedback-driven suppression of star formation.
    \item A significant fraction of blue spirals show high X-ray luminosities, with 40.9\% (9 out of 22) having $L_{X}\geq 10^{41}$ erg s$^{-1}$ and 22.7\% (5 out of 22) exceeding $L_{X}\geq 10^{42}$ erg s$^{-1}$, despite being classified as star-forming or composite (star-forming/AGN) in optical diagrams, indicating hidden or optically obscured AGN activity.
    \item The combined use of BPT, WHAN, and extended diagnostics \citep[e.g.,][]{Mazzolari2024,Mazzolari2025,Backhaus2025} improves AGN identification by recovering additional AGNs in galaxies where one or more classical BPT lines are weak or undetected. For galaxies with most of the emission-line coverage, the classical BPT diagram identifies AGN/LINER activity in $\sim$85\% of red spirals and $\sim$21\% of blue spirals, while extended low-metallicity and low-S/N diagnostics increase the recovered AGN fraction to $\sim$70–100\% in red spirals and $\sim$12–100\% in blue spirals, demonstrating that extended diagnostics are more efficient for identifying AGNs in emission-line-weak and low-metallicity systems, whereas the classical BPT diagram remains robust for strong-line galaxies.
    \item The partial mismatch between optical and X-ray classifications, especially in blue spirals, supports the role of multi-wavelength diagnostics to uncover AGNs missed in traditional optical surveys due to dilution by H II regions.
    \item These results support a dual-mode AGN feedback scenario, where AGNs in red spirals are associated with star formation quenching (negative feedback), while those in blue spirals may contribute to or coexist with ongoing star formation (positive feedback).
\end{itemize}

\section*{Acknowledgments}

This research was partially funded by ND NASA EPSCoR.

This research has made use of the NASA/IPAC Extragalactic Database (NED), which is operated by the Jet Propulsion Laboratory, California Institute of Technology, under contract with the National Aeronautics and Space Administration. This publication makes use of data products from the Wide-field Infrared Survey Explorer \citep{Wright2010}, which is a joint project of the University of California, Los Angeles, and the Jet Propulsion Laboratory/California Institute of Technology, funded by the National Aeronautics and Space Administration. Based on observations made with the NASA Galaxy Evolution Explorer. GALEX is operated for NASA by the California Institute of Technology under NASA contract NAS5-98034. 

This research has made use of data obtained from the Chandra Source Catalog, provided by the Chandra X-ray Center (CXC). Based on observations obtained with XMM-Newton, an ESA science mission with instruments and contributions directly funded by ESA Member States and NASA.

Funding for SDSS-III has been provided by the Alfred P. Sloan Foundation, the Participating Institutions, the National Science Foundation, and the U.S. Department of Energy Office of Science. The SDSS-III website is http://www.sdss3.org/. SDSS-III is managed by the Astrophysical Research Consortium for the Participating Institutions of the SDSS-III Collaboration including the University of Arizona, the Brazilian Participation Group, Brookhaven National Laboratory, University of Cambridge, Carnegie Mellon University, University of Florida, the French Participation Group, the German Participation Group, Harvard University, the Instituto de Astrofisica de Canarias, the Michigan State/Notre Dame/JINA Participation Group, Johns Hopkins University, Lawrence Berkeley National Laboratory, Max Planck Institute for Astrophysics, Max Planck Institute for Extraterrestrial Physics, New Mexico State University, New York University, Ohio State University, Pennsylvania State University, University of Portsmouth, Princeton University, the Spanish Participation Group, University of Tokyo, University of Utah, Vanderbilt University, University of Virginia, University of Washington, and Yale University. This research made use of the ``K-corrections calculator'' service available at\\ http://kcor.sai.msu.ru/. 

Based on observations obtained with MegaPrime/MegaCam, a joint project of CFHT and CEA/IRFU, at the Canada-France-Hawaii Telescope (CFHT), which is operated by the National Research Council (NRC) of Canada, the Institut National des Science de l'Univers of the Centre National de la Recherche Scientifique (CNRS) of France, and the University of Hawaii.

We thank the reviewer for providing thoughtful comments and suggestions which improved the manuscript.

\clearpage

%%%%%%%%%%%%%%%%%%%%%%%%%%%%%%%%%%%%%%%%%%%%%%%%%
\begin{table*}
\normalsize
\centering
\caption{Summary of galaxy classification using optical emission-line ratios.}
\begin{tabular}{|c|c|c|c|c|}\hline
Emission Line Ratio & No. of Galaxies & Star-Forming & AGN/Seyfert & LINERs \\ \hline
log ([O III]/H$_{\beta}$) & Blue (102) & 81 (79.4\%) & 19 (18.6\%) & 2 (2\%) \\ 
vs &&&&\\
log ([N II]/H$_{\alpha}$) & Red (71) & 11 (15.5\%) & 46 (64.8\%) & 14 (19.7\%) \\ \hline
log ([O III]/H$_{\beta}$) & Blue (91) & 85 (93.4\%) & 3 (3.3\%) & 3 (3.3\%)\\
vs &&&&\\
log ([S II]/H$_{\alpha}$) & Red (54) & 15 (27.8\%) & 3 (5.6\%) & 36 (66.7\%)\\ \hline
log ([O III]/H$_{\beta}$) & Blue (31) & 25 (80.6\%) & 4 (12.9\%) & 2 (6.4\%) \\
vs &&&&\\
log ([O I]/H$_{\alpha}$) & Red (42) & 2 (4.8\%) & 5 (11.9\%) & 35 (83.3\%) \\ \hline
log ([O III]/H$_{\gamma}$) & Blue (10) & $0^{*}$ (0\%) & 10 (100\%) & \\
vs &&&&\\
log ([O III]/[O II]) & Red (68) & $2^{*}$ (2.9\%) & 66 (97.1\%) & \\ \hline
log ([O III]/H$_{\gamma}$) & Blue (9) & 1 (11.1\%) & 7 (77.8\%) & $1^{*}$ (11.1\%) \\
vs &&&&\\
log ([Ne III]/[O II]) & Red (56) & 2 (3.6\%) & 37 (66.1\%) & $17^{*}$ (30.3\%) \\ \hline
log ([O III]/[O II]) & Blue (97) & 85 (87.6\%) & 8 (8.2\%) & 4 (4.1\%) \\
vs &&&&\\
log ([N II]/H$_{\alpha}$) & Red (75) & 22 (29.3\%) & 24 (32\%) & 29 (38.7\%) \\ \hline
log ([O III]/H$_{\alpha}$) & Blue (95) & 88 (92.6\%) & 7 (7.4\%) & 0 (0\%) \\
vs &&&&\\
log ([N II]/H$_{\alpha}$) & Red (62) & 15 (24.2\%) & 37 (59.7\%) & 10 (16.1\%) \\ \hline
\end{tabular}
\begin{tablenotes}
\item {$^*$}{Based on \citet{Mazzolari2024} ``Star-forming or AGN/Seyfert'' classification.}
\end{tablenotes}
\label{tab:galaxy_classification}
\end{table*}
%%%%%%%%%%%%%%%%%%%%%%%%%%%%%%%%%%%%%%%%%%%%%%%%%%%%%%

\clearpage

%%%%%%%%%%%%%%%%%%%%
\begin{table*}
\normalsize
\centering
\caption{Properties of galaxies that are located in the AGN section of the {\it WISE} diagram.}
\begin{tabular}{|l|l|l|c|c|l|}\hline
\multicolumn{1}{|c|}{Cluster} & \multicolumn{1}{|c|}{RA (deg)} & Dec (deg) & $L_{\text{x}}$ (erg $\text{s}^{-1}$) & $\Delta L_{\text{x}}$ (erg $\text{s}^{-1}$) & \multicolumn{1}{|c|}{{\it WISE} Classification}\\ \hline
A3158 & 55.4353981  & -53.7059441 & $1.469\times 10^{43}$ & $4.724\times 10^{41}$ & Blue (AGN) \\ \hline
A500  & 69.6615982  & -22.0569439 & $2.912\times 10^{43}$ & $3.143\times 10^{41}$ & Blue (AGN) \\ \hline
A655  & 126.456570   & 47.1745300 & $5.056\times 10^{43}$ & $2.051\times 10^{42}$ & Blue (AGN) \\ \hline
A795  & 141.122570   & 14.3926220   & $5.336\times 10^{42}$ & $1.431\times 10^{42}$ & Blue (SF) \\ \hline
A1291 & 173.268585 & 56.1747971  & $1.312\times 10^{42}$ & $6.713\times 10^{41}$ & Blue (SF) \\ \hline
A1736 & 202.041656 & -27.3318005 & $9.950\times 10^{42}$ & $4.106\times 10^{41}$ & Red (AGN) \\ \hline
A1920 & 216.693880   & 55.2271420 & Not detected & Not detected & Blue (AGN) \\ \hline
A2199 & 247.636030   & 39.3841500    & $2.994\times 10^{42}$ & $6.199\times 10^{40}$ & Blue (AGN) \\ \hline
\end{tabular}
\label{tab:wise_agn}
\end{table*}
%%%%%%%%%%%%%%%%%%%%%%%%%%%%%%%%%%%%%%%%%%%%%%%%%%%%%%%

\clearpage

%%%%%%%%%%%%%%%%%%%%%%%%%%%%%%%%%%%%%%%%%%%%%%%%%%%
\begin{table*}
%\scriptsize
\normalsize
\centering
\caption{K-S test result statistics of red and blue spiral galaxies using emission-line diagnostic diagrams.}
\begin{tabular}{|l|c|c|}\hline
Emission Line Ratio & D-Statistic & Probability\\ \hline
log ([O III]/H$\beta$) vs. log ([N II]/H$\alpha$) & 0.63 & 0.00 \\ \hline
log ([O III]/H$\beta$) vs. log ([S II]/H$\alpha$) & 0.68 & 0.00 \\ \hline
log ([O III]/H$\beta$) vs. log ([O I]/H$\alpha$)    & 0.78 & 0.00 \\ \hline
log ([O III]/H$\gamma$) vs. log ([O III]/[O II]) & 0.31 & 0.40 \\ \hline
log ([O III]/H$\gamma$) vs. log ([Ne III]/[O II]) & 0.19 & 0.50 \\ \hline
log ([O III]/[O II]) vs. log ([N II]/H$\alpha$) & 0.43 & 0.00 \\ \hline
log ([O III]/H$\alpha$) vs. log ([N II]/H$\alpha$) & 0.43 & 0.00 \\ \hline
\end{tabular}
\label{tab:ks_test}
\end{table*}
%%%%%%%%%%%%%%%%%%%%%%%%%%%%%%%%%%%%%%%%%%%%%%%%%%%%%

\clearpage

%%%%%%%%%%%%%%%%%%%%%%%%%%%%%%%%%%%%
\begin{table*}
\scriptsize
\centering
\caption{Classification of blue spirals based on emission-line ratio diagrams.}
\begin{tabular}{|c|c|c|c|c|c|c|c|c|c|}\hline
$\log L_X$& $\Delta L_X$ & [O III]/H$\beta$ & [O III]/H$\beta$ & [O III]/H$\beta$ & [O III]/H$\gamma$ & [O III]/H$\gamma$ & [O III]/[O II] & [O III]/H$\alpha$ & \\
(erg s$^{-1}$) & (erg s$^{-1}$) & vs & vs & vs & vs & vs & vs & vs & WHAN \\
&  & [N II]/H$\alpha$ & [S II]/H$\alpha$ & [O I]/H$\alpha$ & [O III]/[O II] & [Ne III]/[O II] & [N II]/H$\alpha$ & [N II]/H$\alpha$ & \\ \hline
$2.994\times 10^{42}$ & $6.199\times 10^{40}$ & AGN  & AGN  & AGN  & None   & None   & AGN    & AGN   & Retired \\ \hline
$5.909\times 10^{40}$ & $3.668\times 10^{40}$ & SF   & SF   & None & None   & None  & LINERs & LINERs& Retired \\ \hline
$1.214\times 10^{41}$ & $1.644\times 10^{40}$ & AGN  & AGN  & AGN  & None   & None  & AGN    & AGN   & Retired \\ \hline
$2.299\times 10^{41}$ & $1.428\times 10^{40}$ & SF   & SF   & None & None   & None  & None   & SF    & Retired \\ \hline
$1.152\times 10^{41}$ & $2.810\times 10^{40}$ & None & None & None & None   & None  & None   & None  & Retired \\ \hline
$5.056\times 10^{43}$ & $2.051\times 10^{42}$ & AGN  & None & None & None & AGN/SF  & None   & LINERs& None \\ \hline
$5.336\times 10^{42}$ & $1.431\times 10^{42}$ & AGN  & SF   & None & None   & None  & LINERs & LINERs& Retired \\ \hline
$1.312\times 10^{42}$ & $6.713\times 10^{41}$ & AGN  & AGN  & AGN  & None   & None  & SF     & AGN   & None \\ \hline
$3.402\times 10^{40}$ & $2.952\times 10^{40}$ & SF   & SF   & SF   & None   & None  & SF     & SF    & Retired \\ \hline
\end{tabular}
\label{tab:4}
\end{table*}
%%%%%%%%%%%%%%%%%%%%%%%%%%%%%%%%%%%%%%

\clearpage

%%%%%%%%%%%%%%%%%%%%%%%%%%%%%%%%%%%%%%%%
\begin{table*}
\scriptsize
\centering
\caption{Classification of red spirals based on emission-line ratio diagrams.}
\begin{tabular}{|c|c|c|c|c|c|c|c|c|c|}\hline
$\log L_X$ & $\Delta L_X$ & [O III]/H$\beta$ & [O III]/H$\beta$ & [O III]/H$\beta$ & [O III]/H$\gamma$ & [O III]/H$\gamma$ & [O III]/[O II] & [O III]/H$\alpha$ & \\
(erg s$^{-1}$) &  (erg s$^{-1}$) & vs & vs & vs & vs & vs & vs & vs & WHAN \\
&  & [N II]/H$\alpha$ & [S II]/H$\alpha$ & [O I]/H$\alpha$ & [O III]/[O II] & [Ne III]/[O II] & [N II]/H$\alpha$ & [N II]/H$\alpha$ & \\ \hline
$1.701\times 10^{40}$ & $1.243\times 10^{40}$ & None & None & None & AGN     & SF/AGN & LINERs & LINERs & Passive \\ \hline
$1.635\times 10^{40}$ & $3.850\times 10^{40}$ & Comp & SF   & None & None    & None   & None   & None   & Passive \\ \hline
$2.842\times 10^{41}$ & $9.958\times 10^{39}$ & None  & None & None & AGN  & None   & LINERs & LINERs & Passive \\ \hline
$9.917\times 10^{39}$ & $4.749\times 10^{39}$ & None  & None & None & AGN     & SF/AGN & LINERs & LINERs & Passive \\ \hline
$8.226\times 10^{39}$ & $3.045\times 10^{39}$ & SF  & SF   & None & None    & None   & SF     & SF     & Retired \\ \hline
$7.022\times 10^{40}$ & $3.865\times 10^{40}$ & LINERs & None & None & AGN   & AGN   & AGN    & None   & None \\ \hline
$4.624\times 10^{41}$ & $3.078\times 10^{40}$ & None  & None & None & AGN     & SF/AGN & LINERs & LINERs & Passive \\ \hline
$9.511\times 10^{40}$ & $2.253\times 10^{40}$ & AGN/Comp & SF  & None & AGN  & None & AGN & SF & Retired \\\hline
\end{tabular}
\label{tab:5}
\end{table*}
%%%%%%%%%%%%%%%%%%%%%%%%%%%%%%%%%%%%%%%%%

\clearpage

\appendix
\label{sec:appendix}
\setcounter{figure}{0}
\section{Comparison of X-ray flux measurements}

\begin{figure*}
\centering
\includegraphics[scale=1.1]{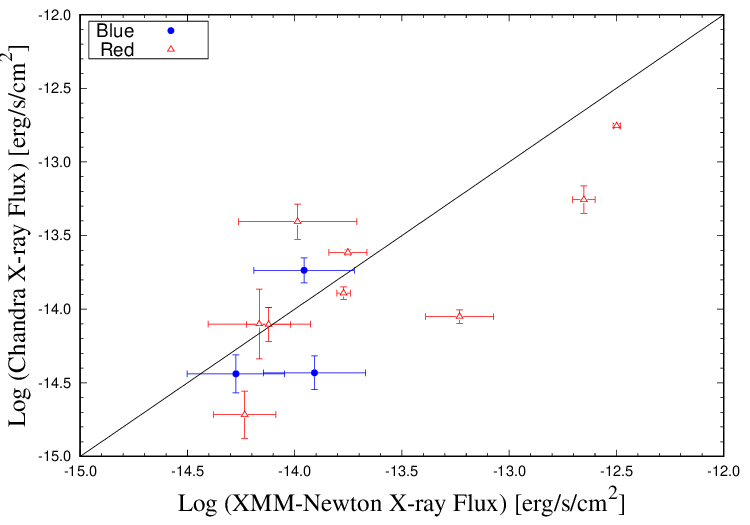}
\caption[Comparison between {\it XMM-Newton} and {\it Chandra} X-ray flux measurements.]{Comparison between {\it XMM-Newton} and {\it Chandra} X-ray flux measurements for three blue spirals and nine red spiral galaxies that have measured fluxes from both telescopes. Two red spirals have flux measurements that disagree by $>3\sigma$. The solid black line depicts equal flux measured by both telescopes.}
\label{fig:A1}
\end{figure*}

In Figure~\ref{fig:A1} we show the comparison of X-ray flux measured for three blue spirals and nine red spirals observed by both {\it Chandra} and {\it XMM-Newton} telescopes. Flux measurements agree to within $3\sigma$ for 10 of the 12 spirals. For two red spirals, the flux differs by $10.2\sigma$ and $5.7\sigma$. 

%%%%%%%%%%%%%%%%%%%%%%
%%%%%%%%%%%%%%%%%%%%%%%%

\end{document}